\documentclass[notitlepage,prd,twocolumn,showpacs,preprintnumbers,nofootinbib,superscriptaddress,floatfix,tightenlines]{revtex4-2}

\usepackage{mathrsfs}
\usepackage{amssymb}
\usepackage{amsmath}
\usepackage{graphicx}
\usepackage{amsfonts,amsbsy}
\usepackage{bm}
\usepackage{nicefrac}
\usepackage[breaklinks,colorlinks,urlcolor=blue,citecolor=citcolor,linkcolor=lcolor]{hyperref}
\usepackage{cleveref}
\usepackage{multirow}
\usepackage{makecell}
\usepackage{xcolor}
\usepackage{booktabs} 
\usepackage{nicematrix}
\usepackage{bigdelim}
\definecolor{lcolor}{rgb}{0.5,0,0}
\definecolor{citcolor}{rgb}{0,0.3,0.0}
\definecolor{ao(english)}{rgb}{0.0, 0.5, 0.0}
\usepackage{comment}

\newcommand{\ns}{n_\mathrm{sat}}

\newcommand{\nPT}{n_\mathrm{PT}}
\newcommand{\ntov}{n_\mathrm{TOV}}

\newcommand{\dn}{\Delta n}

\begin{document}

\title{{First-order phase transitions in the cores of neutron stars}}

\author{Oleg Komoltsev}
\email{oleg.komoltsev@uis.no}
\affiliation{Faculty of Science and Technology, University of Stavanger, 4036 Stavanger, Norway}
\begin{abstract}


I explore various scenarios for the phase transition within neutron-star matter. I do so by generating large model-agnostic ensemble using Gaussian processes, both with and without explicit inclusion of first-order phase transitions (FOPTs). The ensemble is conditioned with state-of-the-art astrophysical and theoretical inputs in a fully Bayesian approach. I study how the current data affect the posterior probability of the location and the strength of FOPT. I find that the previously observed peak structure of the sound speed remains stable against inclusion of FOPTs. While the current data cannot differentiate between a smooth crossover and a first-order phase transition, 91\% of the total evidence consists of equations of state with some form of phase changes, such as FOPT occurring within or terminating the stable branch of neutron stars, or an indication of a crossover to quark matter.

\end{abstract}

\maketitle
\section{Introduction}

The phase structure of matter, when compressed to several times nuclear saturation density ($\ns\approx 0.16$ fm$^3$), is one of the central questions in the study of neutron stars (NSs). While it remains unclear how to theoretically probe cold dense matter, recent progress in astrophysical observations of NSs has given rise to rapid development in the study of such matter \cite{Kurkela:2014vha,Most:2018hfd,Annala:2017llu, Tews:2018iwm,Landry:2018prl,Capano:2019eae,Miller:2019nzo,Essick:2019ldf,Raaijmakers:2019dks,Annala:2019puf,Dietrich:2020efo,Landry:2020vaw,Al-Mamun:2020vzu,Miller:2021qha,Raaijmakers:2021uju,
Annala:2021gom,Huth:2021bsp,Altiparmak:2022bke,Lim:2022fap,Gorda:2022jvk,Annala:2023cwx,Komoltsev:2023zor,Koehn:2024set, Ecker:2024eyf, Fujimoto:2024cyv, Fujimoto:2022ohj, PhysRevD.105.023018, Alford:2015gna}. This work is among those studies.

The structure of NSs is determined by the equation of state (EoS), which, for strongly interacting matter, is governed by Quantum Chromodynamics (QCD). In practice, \textit{ab-initio} theoretical calculations can only be performed in two limits. The low-density part, corresponding to hadronic matter, can be constrained by Chiral Effective Field Theory (cEFT) up to [1.1-2]$\ns$ \cite{Hebeler:2013nza,Drischler:2017wtt,Drischler:2020hwi,Tews:2018kmu, PhysRevLett.130.072701}. On the other hand, the perturbative QCD calculations (pQCD) \cite{Kurkela:2009gj,Kurkela:2016was,Gorda:2018gpy,Gorda:2021znl,Gorda:2023mkk, Gorda:2021gha} at high-density regime inform us about the behavior of quark matter, which are reliable starting from [25-40]$\ns$ \cite{Gorda:2023usm}. The densities reached within the most massive neutron stars are around 3-8 $\ns$, where the matter is on the verge of dissolving into deconfined quarks \cite{Annala:2019puf, Annala:2023cwx}. It remains an open question of whether the phase transition (PT) from hadronic to quark matter occurs in the cores of NSs - the densest matter found in the Universe.

The second fundamental question that arises is how the phase transition happens. Some alternative scenarios include the smooth crossover leading to quark matter cores, a violent FOPT with a discontinuous change in density destabilizing NSs and leading to collapse into a black hole (BH), or a more exotic phase transition suggesting a new state of matter. However, as of yet, there is no theoretical basis to favor any one of these scenarios. 

A natural approach to addressing these questions is by exploiting a model-agnostic method to generate large ensemble of different EoSs. By conditioning the ensemble with astrophysical observations and \textit{ab-initio} theoretical calculations, we can estimate evidences for different scenarios of the phase change. There have been multiple inferences of the EoS \cite{Kurkela:2014vha,Landry:2018prl,Miller:2019nzo,Essick:2019ldf,Raaijmakers:2019dks,Annala:2019puf,Dietrich:2020efo,Landry:2020vaw,Al-Mamun:2020vzu,Annala:2021gom,Altiparmak:2022bke,Gorda:2022jvk,Annala:2023cwx,Komoltsev:2023zor,Koehn:2024set}, most of which do not include a first-order phase transition. While there are studies examining the first-order PT in neutron-star matter, they have certain limitations. Some works do not explicitly model FOPT \cite{Brandes:2023hma, Essick:2023fso}, while others lack a Bayesian interpretation \cite{Gorda:2022lsk, PhysRevC.107.025801, PhysRevD.88.083013}, or utilize a restrictive prior \cite{Mroczek:2023zxo}. As of yet, no studies utilize the recent development of QCD input \cite{Komoltsev:2023zor}, and to the best of my knowledge, there is no Bayesian inference of the FOPT properties.

In this Letter, I perform a state-of-the-art fully Bayesian inference of the EOS using Gaussian processes (GP) regression \cite{Landry:2018prl, Landry:2020vaw, Gorda:2022jvk, Mroczek:2023zxo} with explicit inclusion of the first-order phase transition. The astrophysical inputs include tidal-deformability (TD) measurements from the GW170817 \cite{TheLIGOScientific:2017qsa,LIGOScientific:2018cki,LIGOScientific:2018hze}, the hypothesis that the remnant in GW170817 collapsed to a BH \cite{Margalit:2017dij,Rezzolla:2017aly,Ruiz:2017due,Shibata:2017xdx,Shibata:2019ctb}, 
radio mass measurement of the heaviest pulsar \cite{Antoniadis:2013pzd}, and 12 X-ray measurements of mass and radius of NS \cite{Miller:2019cac,Riley:2019yda,Fonseca:2021wxt,Miller:2021qha,Riley:2021pdl,Shaw:2018wxh,Steiner:2017vmg,Nattila:2017wtj,Nattila:2015jra}. The theoretical constraints come from the low-density regime provided by cEFT up to 2$\ns$ \cite{Hebeler:2013nza,Drischler:2017wtt,Drischler:2020hwi,Tews:2018kmu} and high-density pQCD calculations using either conservative QCD input \cite{Komoltsev:2021jzg, Gorda:2022jvk} or the marginalized QCD likelihood function \cite{Komoltsev:2023zor}.

I categorize each generated EoS into one of several mutually exclusive sets, each representing the most interesting scenario:
\begin{description}
    \item[No FOPT] there is no first-order PT in the stable neutron star branch. 
    \item[FOPT inside NS] there is a first-order PT in the stable neutron star branch. The NS remains stable after FOPT. 
    \item[Destabilizing FOPT] there is a first-order PT in the stable neutron star branch, after which NS collapses into a BH. 
    \item[Twin stars] there is a second stable branch in mass-radius curve, irrespective of the location of the FOPT. 
\end{description}

The main objective of the study is to compare evidences of each set based on the current data. Furthermore, I evaluate how the explicit inclusion of the FOPT into the prior changes the previously determined probability of having quark matter cores inside the most massive NSs \cite{Annala:2023cwx}. 
In the case of a first-order PT, I calculate the posterior distribution for the beginning $\nPT$ and the strength $\dn$ of PT. 
Lastly, I examine how future mass-radius measurements of NS affect the evidence of the sets. This analysis pinpoints which possible observations could help in distinguishing between various potential scenarios for the phase transition.

\begin{figure*}[ht!]
    \centering
\includegraphics[width=0.85\textwidth]{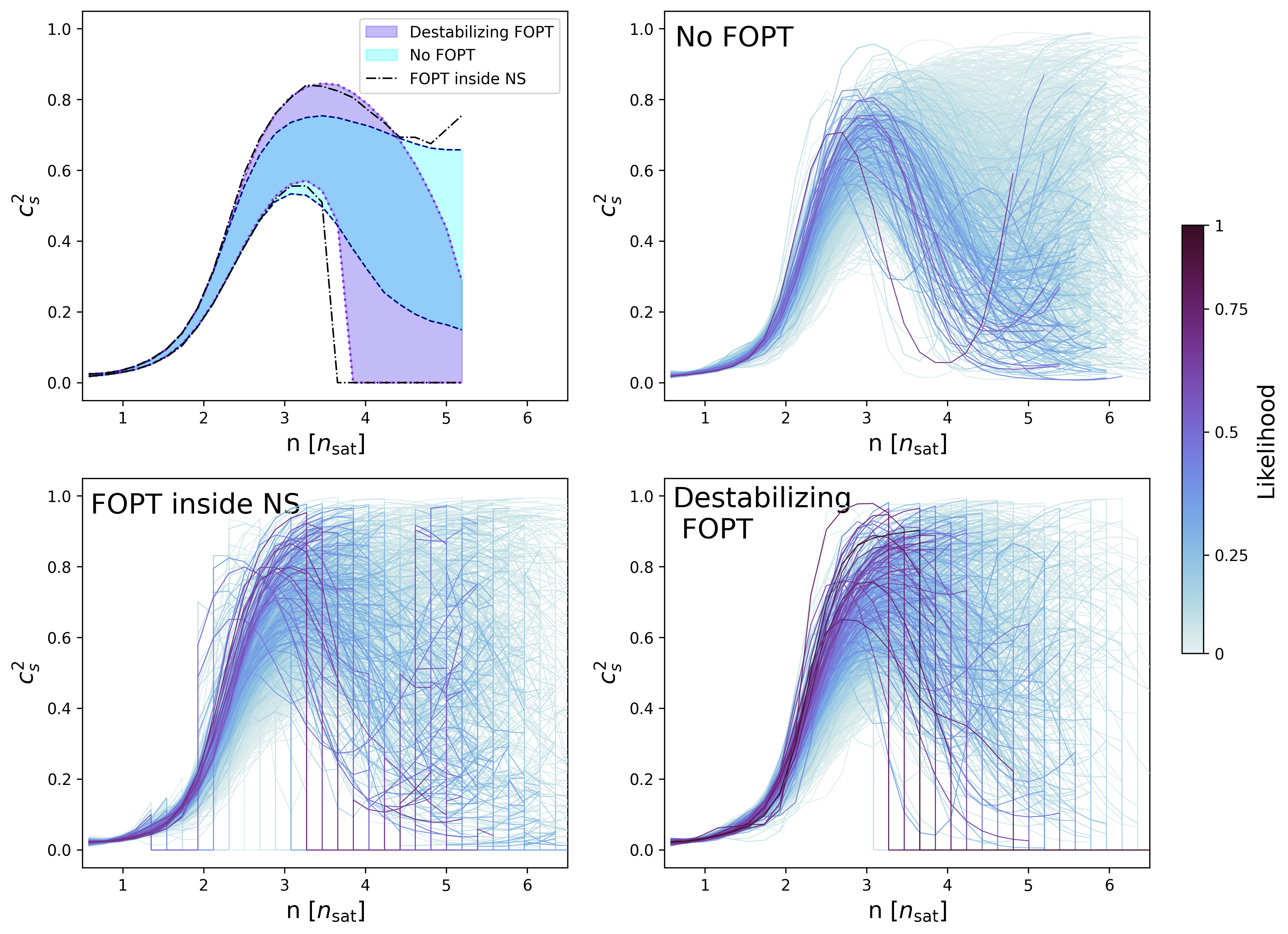}
    \caption{(\textbf{Upper left}) Sixty-eight percent CI for $c^2_s$ for different sets. An EoS contributes to the CI only up to TOV density. The plot shows the conditional posterior distribution $\mathrm{P}(c^2_s|n, n<\ntov)$ up to fixed $n$. (\textbf{Other panels}) A representative sample of EoSs for each set. The color coding of each EoS corresponds to the likelihood assigned according to  \cref{eq:likelihoods}, normalized to the maximum likelihood in the ensemble.}
    \label{cs2}
\end{figure*}

\section{Setup}

In this section, I present a brief summary of the prior choices, theoretical and astrophysical inputs that have been used. GP regression is performed in an auxiliary variable $\phi(n)=-\ln (1/c_s^2 - 1)$, with the prior distribution being a multivariate Gaussian distribution:
\begin{equation}
\phi(n) \sim \mathcal{N}\bigl(-\ln (1/\bar{c}_s^2 - 1), K(n, n')\bigr)\,,
\end{equation}
where $K(n, n') = \eta \exp \bigl( - (n - n')^2 / 2 \ell ^2 \bigr)$ is a Gaussian kernel. The hyperparameters $\eta$, $\ell$, $\bar{c}_s^2$ are independently drawn for each EoS from the distribution (see Eq. 6 in \cite{Gorda:2022jvk}). The grid spacing in $n$ is around 0.2$\ns$. The details regarding EoS construction using GP regression in $\phi(n)$ can be found in \cite{Gorda:2022jvk}. 

The EoS is constructed from two samples of GP, with a segment of zero sound speed in between, introducing a discontinuous jump in density. Both samples are drawn from the GP conditioned with cEFT up to 1.1$\ns$\footnote{For no particular reason, the cEFT input is divided into two parts: first, GP is conditioned with cEFT up to 1.1$\ns$, and then the cEFT likelihood function is used in the range [1.1, 2]$\ns$}. The hyperparameters, such as the correlating length $\ell$ and the variance $\eta$ of the Gaussian kernel, are identical before and after FOPT. The algorithm can be written in the following way: 

\begin{itemize}
    \item[1.] Select the location of the FOPT from a uniform distribution between 1.1$\ns$ and 10$\ns$, $\nPT \in [1.1,10]\ns$.  
    \item[2.] Draw the strength of FOPT from uniform distribution $\dn \in [0,8]\ns$.
    \item[3.] Draw two independent samples of EoS from the conditioned GP regression.  One sample extends from $1.1\ns$ to $\nPT$, while the second sample spans from $\nPT+\dn$ to $10\ns$.
    \item[4.] Combine these samples to form a single EoS across the range $[1.1,10]\ns$, assuming $c^2_s=0$ within the interval $[\nPT,\nPT+\dn]$. 
\end{itemize}

This procedure ensures that the prior is uniform in the $\nPT$--$\dn$ --plane, and that there is no correlation in $c^2_s$ before and after FOPT. 

Subsequently, I solve the Tolman-Volkoff-Oppenheimer (TOV) equation to predict mass-radius relation \cite{Tolman:1939jz, Oppenheimer:1939ne} as well as determine the maximal density $\ntov$ that is supported by the stable nonrotating NS. To capture the behavior of tidal deformability (TD) with discontinuous jump in number density in the case of first-order PT I follow the approach outlined in  \cite{Han:2018mtj}. Each EoS is used only up to TOV-density; thus, densities greater than $\ntov$ are excluded from the analysis.

Following the approach outlined in \cite{Tews:2018kmu} and reviewed in \cite{Koehn:2024set} to incorporate more information from the low-density calculations, I employ the cEFT likelihood function within the interval [1.1,2]$\ns$ (as provided by eqs. (4) and (5), and fig. 2 in \cite{Koehn:2024set}). To obtain a likelihood for EoS, the function is integrated over the interval [1.1,2]$\ns$, penalizing EoSs that fall outside the range [$p_+, p_-$] provided in \cite{Koehn:2024set}.

Two types of QCD inputs are considered. Conservative QCD input was introduced in \cite{Komoltsev:2021jzg} and \cite{Gorda:2022jvk} (publicly available in \cite{komoltsev_oleg_2023_7781233}). It verifies whether the density up to which an EoS is modeled (chosen to be $\ntov$ in this study) can be connected by at least one causal, stable, and consistent EoS to the perturbative QCD limit at 40 $\ns$. The marginalized QCD likelihood function, introduced in \cite{Komoltsev:2023zor} and publicly available in \cite{komoltsev_2025_15407795}, incorporates additional information on the well-convergent sound speed in the perturbative regime starting from 25$\ns$. It is marginalized over another GP prior in the range [$\ntov,40\ns$]. This GP prior is conditioned with pQCD calculation of $c^2_s$ within [25,40]$\ns$ and does not have first-order PT. Both conservative and marginalized QCD inputs are imposed at TOV density.

\begin{table}[ht!]
\caption{A summary of all mass and mass-radius measurements that have been used to condition the ensemble.}
\label{table:astro}
\begin{NiceTabular}{lc}[code-before = \rowcolor{blue!15}{2,4,7,15}]
\toprule
\textbf{Name} & \textbf{Ref.}\\
\addlinespace[0.3em]
\multicolumn{2}{c}{Radio measurement} \\
\addlinespace[0.3em]
PSR J0348+0432 & \cite{Antoniadis:2013pzd}\\
\addlinespace[0.3em]
\multicolumn{2}{c}{NICER pulsars} \\
\addlinespace[0.3em]

PSR J0030+0451 &\cite{Miller:2019cac,Riley:2019yda} \\
PSR J0740+6620 & \cite{Fonseca:2021wxt,Miller:2021qha,Riley:2021pdl}\\
\addlinespace[0.3em]
\multicolumn{2}{c}{qLMXB systems} \\
\addlinespace[0.3em]

M13& \cite{Shaw:2018wxh}\\
M28& \cite{Steiner:2017vmg}\\
M30& \cite{Steiner:2017vmg}\\
$\omega$ Cen& \cite{Steiner:2017vmg}\\
NGC 6304 & \cite{Steiner:2017vmg}\\
NGC 6397& \cite{Steiner:2017vmg}\\
47 Tuc X7 & \cite{Steiner:2017vmg}\\
\addlinespace[0.3em]
\multicolumn{2}{c}{X-ray bursters} \\
\addlinespace[0.3em]
4U 1702-429 & \cite{Nattila:2017wtj}\\
4U 1724-307& \cite{Nattila:2015jra} \\
SAX J1810.8-260& \cite{Nattila:2015jra}\\
\bottomrule

\end{NiceTabular}
\end{table}

The ensemble is conditioned with astrophysical inputs identical to those in \cite{Annala:2023cwx}. The methodology is elaborated in \cite{Gorda:2022jvk} and \cite{Annala:2023cwx}. Mass and mass-radius measurements are summarized in \cref{table:astro}. For gravitational-wave (GW) data, I use the joint posterior density function for the TD parameter $\Tilde{\Lambda}$ and mass ratio at fixed chirp mass from GW170817 by the LIGO-Virgo Collaboration \cite{TheLIGOScientific:2017qsa,LIGOScientific:2018cki,LIGOScientific:2018hze}. Additionally, as suggested by the electromagnetic counterpart of GW170817 \cite{LIGOScientific:2017ync}, the resulting binary merger product is a BH \cite{Margalit:2017dij,Rezzolla:2017aly,Ruiz:2017due,Shibata:2017xdx,Shibata:2019ctb}, which puts constraints on the total baryon mass that can be supported by the EoS.

All the inputs are subsequently incorporated into Bayes' theorem: 
\begin{equation}
P({\rm EoS }\,|\, {\rm data} ) = \frac{ P({\rm EoS}) \, P(  {\rm data} \,|\, {\rm EoS} \,)}{P({\rm data})},
\label{eq:bayes}
\end{equation}
where $P({\rm data} \,|\,  \rm EoS)$ is the product of uncorrelated likelihoods, which can be written as
\begin{align}
P({\rm data} & \,|\, {\rm EoS} ) =  P({\rm QCD} \,|\, {\rm EoS})  P({\rm cEFT} \,|\, \rm EoS)\nonumber \\
 \times &P( {\rm X\text{-}rays} \,|\, {\rm EoS}) P({\rm \tilde \Lambda}, {\rm BH}  \,|\, {\rm EoS}) P({\rm Radio}  \,|\, \rm EoS).
 \label{eq:likelihoods}
\end{align}
These likelihoods correspond to the QCD and cEFT inputs, product of likelihoods for all x-rays measurements listed in \cref{table:astro} (under NICER pulsars, qLMXB systems and x-ray bursters), along with constraints from GW170817 and radio mass measurement. All the plots in this Letter are produced using all the mentioned astrophysical data, cEFT likelihood function up to 2$\ns$ as well as marginalized QCD input (cf. \cref{fig:cs2_conserv} in the supplemental material).

The ensemble is then divided into the sets described in the introduction, based on the location of the FOPT with respect to the stable branch of NS. \textit{No FOPT} -- set is constructed from the following EoSs: firstly, if an EoS has $\dn=0$, which happens when $\dn$ is smaller than half of the grid spacing $\sim 0.1\ns$; or secondly, if a phase transition occurs outside the first stable branch, $\ntov<\nPT$, as these densities are excluded and have no effect on the density below TOV. An EoS is labeled to have \textit{destabilizing FOPT} if the FOPT happens in the stable NS branch and the first grid point after FOPT falls in the unstable branch. In this case the TOV density is taken to be the last grid point of the FOPT: $\ntov=\nPT+\dn$. \textit{FOPT inside NS} is identified if $\nPT+\dn < \ntov$. An EoS is assigned to \textit{twin stars} -- set if it has second stable branch, irrespective of the FOPT location. For twins stars the astrophysical likelihoods are marginalized over both stable branches, and QCD input is imposed at the maximal density of the second branch.

The total number of EoSs in the ensemble is 300k, distributed among the categories as follows: 66k - \textit{no FOPT}, 37k - \textit{FOPT inside NS}, 121k - \textit{destabilizing FOPT} and 76k - \textit{twin stars}.  

\section{Results}
\subsection{EoS inference}

The samples of the EoSs and the corresponding credible intervals (CI) for speed of sound can be found in \cref{cs2}. Regardless of the presence of a FOPT, the overall structure of the EoS remains similar, characterized by a peak in the sound speed around 2-3$\ns$, followed by a softening. \textit{No FOPT} -- set reproduces the results seen in \cite{Gorda:2022jvk, Altiparmak:2022bke}. The absence of \textit{twins} -- set in the figure is due to their vanishing evidence, as will be explained in the next section.

The peak in sound speed is determined by the mass constraints and becomes more prominent with the inclusion of X-ray measurements. Data from GW170817 and cEFT constraints soften an EoS at lower density. After the peak is reached, all astrophysical constraints are relaxed. Note that rapid stiffening can be explained within, e.g. Quarkyonic matter
model \cite{McLerran:2018hbz}. The behavior of an EoS is solely determined by the QCD input above the peak in the interval [$2M_\odot,M_{\rm TOV}$], which forces an EoS to soften. The forced softening can be either the rapid softening indicating a phase transition or a slower crossover to a conformal value $c^2_s=1/3$ or below. 

The peak tends to be higher for scenarios involving a first-order PT, as also evident from the CI (upper left subplot). Marginalized QCD input utilizes more information from the pQCD calculations and requires a strong softening. The more dramatic the softening, the stiffer an EoS can become before being penalized by QCD constraints, as explained in \cite{Komoltsev:2023zor}. An EoS cannot achieve sound speeds close to unity without experiencing sharp softening immediately after the peak, effectively exhibiting FOPT-like behavior. Consequently, an EoS with a FOPT can reach a higher values of $c^2_s$.

Note that an EoS contributes to CI only up to TOV density. The posterior distribution can only be shown up to the lowest TOV density within the ensemble. However, since destabilizing FOPT tends to have smaller TOV densities, and such a plot would offer limited information. Therefore, a conditional posterior distribution $\mathrm{P}(c_s^2 \mid n, n < n_{\mathrm{TOV}})$ is displayed in the upper left panel of \cref{cs2}.

\subsection{Bayes factors}
\label{sec:bayes}
The Bayes factor quantifies how effectively a model explains the data compared to another model.
Sets, each representing distinct scenarios, can be compared as competing statistical models. The Bayes factor is the ratio of marginalized likelihood, also known as evidence, and can be expressed as
\begin{equation}
\label{bayes_factor}
B^{\rm set_1}_{\rm set_2} = \frac{P(  {\rm set_1} \,|\, \rm data \,)}{P(  {\rm set_2} \,|\, \rm data \,)}\frac{P({\rm set_2})}{P({\rm set_1})},
\end{equation}
where $P(  {\rm set_i} \,|\, \rm data \,)$ represents posterior probability of set $i$ given by \cref{eq:bayes}, and $P({\rm set_i})$ denotes the prior probability of set $i$. Assuming that \textit{a priori} all sets are equally probable, the prior probability simplifies to the number of EoSs in the set. 

\begin{table}[ht]
\caption{A summary of the Bayes factors for various scenarios. Each entry corresponds to the ratio of evidence for a given set to that of the \textit{no FOPT} -- set as defined in \cref{bayes_factor}. The evidences are computed for priors conditioned with radio measurements, cEFT likelihood function and GW data, along with two likelihoods listed in the first column.}
\label{table:1}
\resizebox{0.5\textwidth}{!}{
\begin{NiceTabular}{c|c|c|c}[code-before = \rowcolor{blue!15}{2,4}]
\toprule
$\mathbf{B^{\rm set}_{\mathrm{noFOPT}}}$ & \textbf{Destab. FOPT} & \textbf{FOPT inside NS} & \textbf{Twins} \\
\addlinespace[0.3em]

\makecell{Marginalized QCD\\ $\ \ \ \ \ \ \ $+ X-rays} & 1.5 & 0.7 & 0.001\\ 
\addlinespace[0.3em]
\makecell{Marginalized QCD\\ $\ \ \ \ \ \ \ $+ PSR J0740}& 1.5 & 1.0 & 0.001\\ 
\addlinespace[0.3em]

\makecell{Conservative QCD\\ $\ \ \ \ \ \ \ $+ X-rays} & 0.8 & 0.5 & 0.001 \\
\addlinespace[0.3em]

\makecell{Conservative QCD\\ $\ \ \ \ \ \ \ $+ PSR J0740}& 0.8 & 0.7 & 0.001 \\

\bottomrule
\end{NiceTabular}
}
\end{table}

The Bayes factors obtained in the study are summarized in \cref{table:1}. Each entry represents the ratio of the evidences of the named set in the column to that of the \textit{no FOPT} -- set. For a robust evidence that the data favors set $i$ over the scenario without any FOPT, the Bayes factor must be several times greater than unity (or smaller if the \textit{no FOPT} -- set is preferred). All of the evidences are calculated using radio measurements and GW data constraints. Additionally, either conservative or marginalized QCD input is imposed, as well as two options for x-ray measurements are considered: either all available measurements or solely the NICER measurement of PSR J0740+662. 

By examining the \cref{table:1} the only decisive factor corresponds to \textit{twin stars}, indicating that twin stars are not supported by the current data as $B^{\rm twins}_{\mathrm{noFOPT}}\ll1$. The rest of the Bayes factors are not of any statistical significance, indicating that the current data cannot differentiate between the sets. I speculate that the reason why is that the QCD input, which becomes important in determining the EoS above $2M_\odot$, can be fully satisfied by either FOPT or a crossover. Consequently, there is currently no measurement or theoretical input favoring either scenario. 

Nonetheless, it is interesting to observe how conditioning the prior with different inputs can impact the outcome. The mild increase in the Bayes factor resulting from the change from conservative to marginalized QCD input can be explained by the exclusion of stiff EoSs from the \textit{no FOPT} -- set, thus slightly decreasing the evidence. The choice of x-ray data only affects the intermediate densities below $n < 3n_{\rm sat}$ (see \cref{fig:cs2_conserv} in the supplemental material) and, therefore, it has a limited impact on whether EoS terminates with a crossover or a destabilizing FOPT.

\begin{figure}[h!]
    \centering
\includegraphics[width=0.5\textwidth]{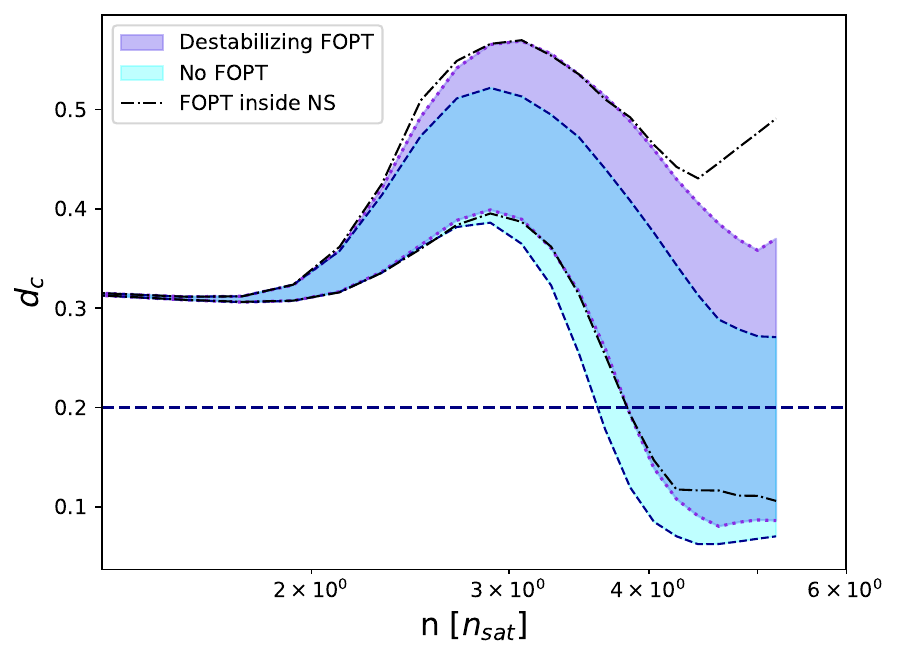}
    \caption{68\%-CI for the $d_c$ parameter for different sets. An EoS contributes to the CI only up to TOV density. The plot shows the conditional posterior distribution $\mathrm{P}(d_c|n, n<\ntov)$ up to fixed $n$.}
    \label{fig:dc}
\end{figure}

It has been suggested in \cite{Annala:2023cwx} that quark matter (QM) can be identified by observing an approximate restoration of conformal symmetry, with the quantity $d_c$ proposed for this purpose:
\begin{eqnarray}
d_{\mathrm{c}}&\equiv&\sqrt{\Delta^2+(\Delta')^2},
\label{eq:paramrels}
\end{eqnarray}
where
\begin{eqnarray}
\Delta &=& \frac{1}{3} - \frac{c_\mathrm{s}^2}{\gamma}\, , \quad
\Delta'  \;=\; c_\mathrm{s}^2 \biggl( \frac{1}{\gamma} - 1 \biggr)\, ,\quad \gamma= \frac{\mathrm{d} \ln p}{\mathrm{d} \ln \varepsilon}.
\end{eqnarray}

68\%-CI for the $d_c$ parameter is shown in \cref{fig:dc}. The posterior probability of a \textit{crossover to QM}, defined as $d_c < 0.2$, in the \textit{no FOPT} set is 64\% for the most massive NSs (cf. 75\% reported in \cite{Annala:2023cwx}). However, with the explicit inclusion of FOPT in the prior, this probability decreases to 50\% for \textit{FOPT inside NSs} set and to 30\% for \textit{destabilizing FOPT} set. While there are EoSs with FOPT inside NSs that result in QM appearing above PT, most EoSs with destabilizing FOPT do not meet this criterion. Only a small portion exhibit a crossover to QM before destabilizing PT. What is particularly intriguing is that softening of the EoS seems to be inevitable, as evident from \cref{cs2}, and thus the peaked structure of the sound speed is stable against inclusion of FOPT in the prior.

The probability of the absence of the phase change can be assessed by comparing the combined sets of \textit{FOPT inside NS}, \textit{destabilizing FOPT}, and EoSs with crossover to QM in \textit{no FOPT} set against the rest of the prior, namely set of EoSs without a crossover to QM in the \textit{no FOPT} set (twins carry vanishing weight). The combined sets make up 91\% of the total evidence, leaving only 9\% for the EoSs without FOPT or crossover to QM. It's important to note that EoSs in the \textit{no FOPT} set that do not meet criteria for QM may mimic FOPT-like behavior. Consequently, introducing any criteria for such "pseudo" first-order PTs would only decrease the 9\% probability further.

\subsection{FOPT posterior}
\label{sec:4}

\begin{figure}[h!]
    \centering
\includegraphics[width=0.5\textwidth]{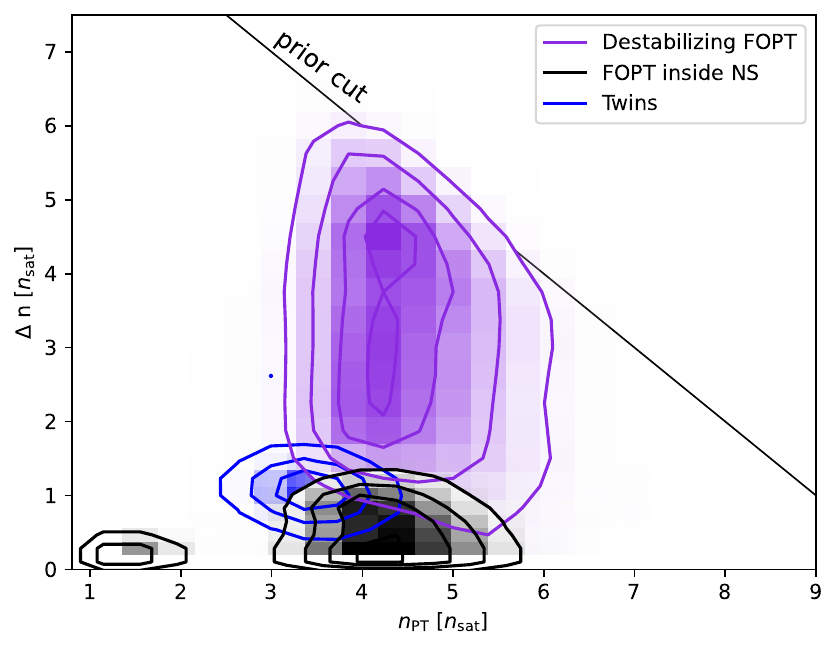}
    \caption{The posterior distribution of the location $\nPT$ and strength $\dn$ of the FOPT. The prior is uniform in the $\nPT$--$\dn$ -- plane. A Gaussian filter is applied to smooth the data.}
    \label{ndn}
\end{figure}

In this section, I present the results of Bayesian inference of the FOPT properties, such as its location and strength. The impact of the astrophysical data and theoretical inputs on the properties of first-order PT is summarized in \cref{ndn}, illustrating the posterior distribution of $\nPT$--$\dn$. The prior distribution is uniform in the $\nPT$--$\dn$ -- plane. Any values $\nPT+\dn$ exceeding $10\ns$ are discarded, as indicated by the prior cut on the figure.

\textit{FOPT inside NS} -- set is shown in black in \cref{ndn}. A notable feature of the posterior distribution is the gap between FOPTs occurring early in the density range [1,2]$\ns$ and the latter PTs within [3,6]$\ns$. The gap in densities corresponds roughly to a mass range spanning from 0.5 to 1.9 $M_\odot$. The existence of the gap is a consequence of the mass constraints, which require that an EoS remains stiff in this interval, mostly excluding FOPTs in this region. Another noteworthy aspect is that any FOPT with $\dn \gtrsim 1.2\ns$ results in the destabilization of NS and cannot support FOPT inside the NS. Note that any small FOPTs do not change an EoS at the level of integral quantities such as pressure and energy density, and therefore no astrophysical data can disfavor an EoS with small FOPT. 

\textit{Destabilizing FOPT} appears only above 3$\ns$, which is the typical minimal density at which an EoS satisfies mass constraints. In principle, if an EoS becomes destabilized due to a FOPT of some strength $\dn$, then any larger FOPT would also lead to the collapse of NS, resulting in a uniform distribution of $\dn$ above the lower bound. However, large FOPTs are slightly penalized by marginalized QCD input, resulting in a peak distribution around $\nPT=4\ns$ and $\dn$ ranging from 2 to 3$\ns$. Twin stars are mostly overlapping with both sets. In addition, twin star solutions allow for some FOPT in the gap below 3$\ns$, although with small posterior weight.

\section{Conclusions}

The results of this Letter demonstrate that the previously discovered peak structure of the sound speed inside NSs remains stable even with the explicit inclusion of the first-order PT in the prior. While the current data constrain the location and the strength of the potential first-order PT, it does not definitively favor any particular scenario. The possibilities, including a small first-order PT within NS, a crossover to quark matter, or a destabilizing FOPT leading to a collapse, remain viable. By generating a broad prior, I evaluate that the probability of the scenario falling outside these categories is low. Most of the total evidence, approximately 91\%, consists of EoSs either featuring FOPT occurring within or terminating the stable branch of NSs, or a crossover to quark matter. This leaves 9\% probability for the absence of any form of the phase change. Moreover, identifying EoSs in \textit{no FOPT} set that mimic FOPT-like behavior would only reduce 9\% probability further. This suggests that nontrivial phase changes can be explored within the densities found in neutron stars.

The challenges in distinguishing between sets are outlined in \cref{sec:bayes}. In brief, the only constraint that propagates within the interval between 2$M_\odot$ and $M_{\rm TOV}$ is the QCD input, which necessitates softening but remains indifferent as to whether an EoS exhibits FOPT-like behavior or conformalization. However, by imposing various mass-radius measurements, we can predict what future detection might help differentiate between scenarios.

\begin{figure}[h!]
    \centering
\includegraphics[width=0.5\textwidth]{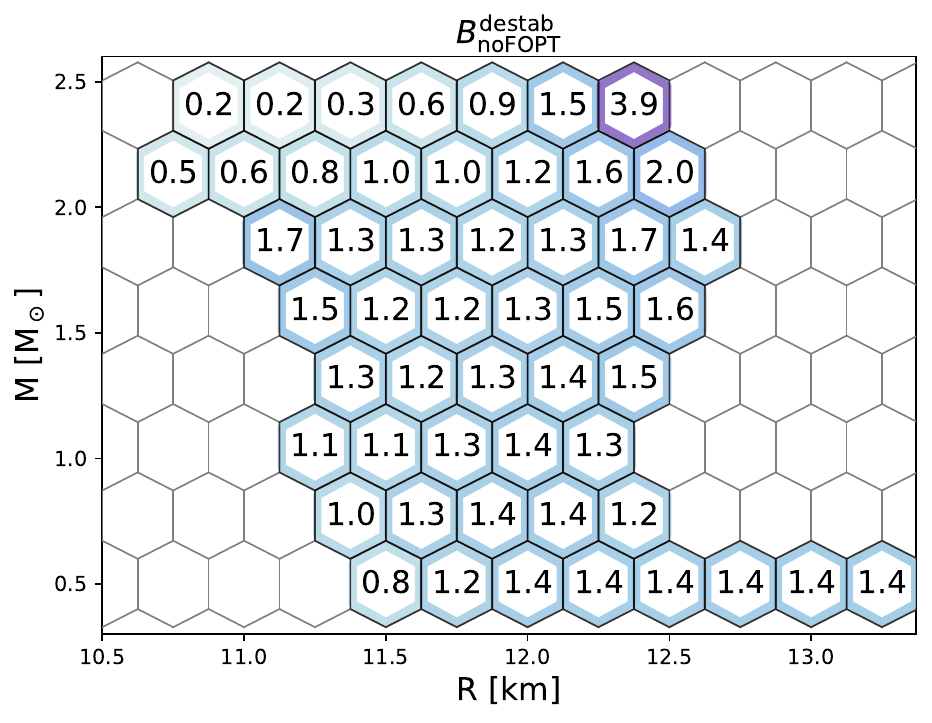}
    \caption{The collection of the Bayes factors for two sets \textit{destabilizing FOPT} and \textit{no FOPT} for future mass-radius measurement. Each hexagon represents a mass-radius measurement, with the likelihood being a step function set to one inside the hexagon and zero outside. The Bayes factor greater than unity indicates data preference toward \textit{destabilizing FOPT}. None of these factors are of any statistical significance.}
    \label{fig:bayes}
\end{figure}

The summary of various Bayes factors obtained by considering potential future $M$-$R$ measurements is presented in \cref{fig:bayes}. Each hexagon represents a mass-radius measurement, wherein the likelihood is assigned a value of one if EoS intersects the hexagon and zero otherwise. To avoid numerical instabilities, only hexagons through which at least 100 EoSs pass with a nonzero weight are taken into account. The Bayes factors are computed to compare the \textit{destabilizing FOPT} set and the \textit{no FOPT} set. Although none of the factors yield decisive conclusions, measurements with high mass and high radius tend to slightly favor the \textit{destabilizing FOPT} set. Note the that effect of black-widow pulsar PSR J0952-0607 with the mass of 2.35$\pm 0.17 M_\odot$ \cite{Romani:2022jhd} depends on the radius of the pulsar. A possible candidate for a light NS, HESS J1731-347 ($M=0.77^{+0.20}_{-0.17} M_\odot$ and $R=10.4^{+0.86}_{-0.78}$km) \cite{2022NatAs...6.1444D}, has a vanishing Bayes factor in the analysis, approximately $10^{-3}$.

As shown in \cref{fig:bayes}, a single mass-radius measurement is insufficient to distinguish between the sets. Low-energy nuclear experiments and tidal deformability measurements can further constrain EoS at intermediate densities. However, to address the question of whether the stable branch of neutron stars terminates with a crossover or FOPT, precise mapping of the EoS will be needed. Such precise mapping can be achieved, i.e by collecting GW data from a large number of binary NS mergers. Yet, even with extensive data, mapping the region near the TOV limit might pose challenges, necessitating stars to be near their maximum mass to constrain this area effectively. Therefore, the interplay between astrophysical constrains and QCD input might be crucial to determine what is happening inside neutron stars. Another potential source of information is the post-merger GW signal from NS mergers \cite{Fujimoto:2022xhv, Ecker:2024eyf}, which can be achieved with the next generation of GW detectors \cite{Evans:2021gyd,Punturo:2010zza}.

\section{Supplemental material }
\label{appendix}
\begin{figure}[ht!]
    \centering
\includegraphics[width=0.5\textwidth]{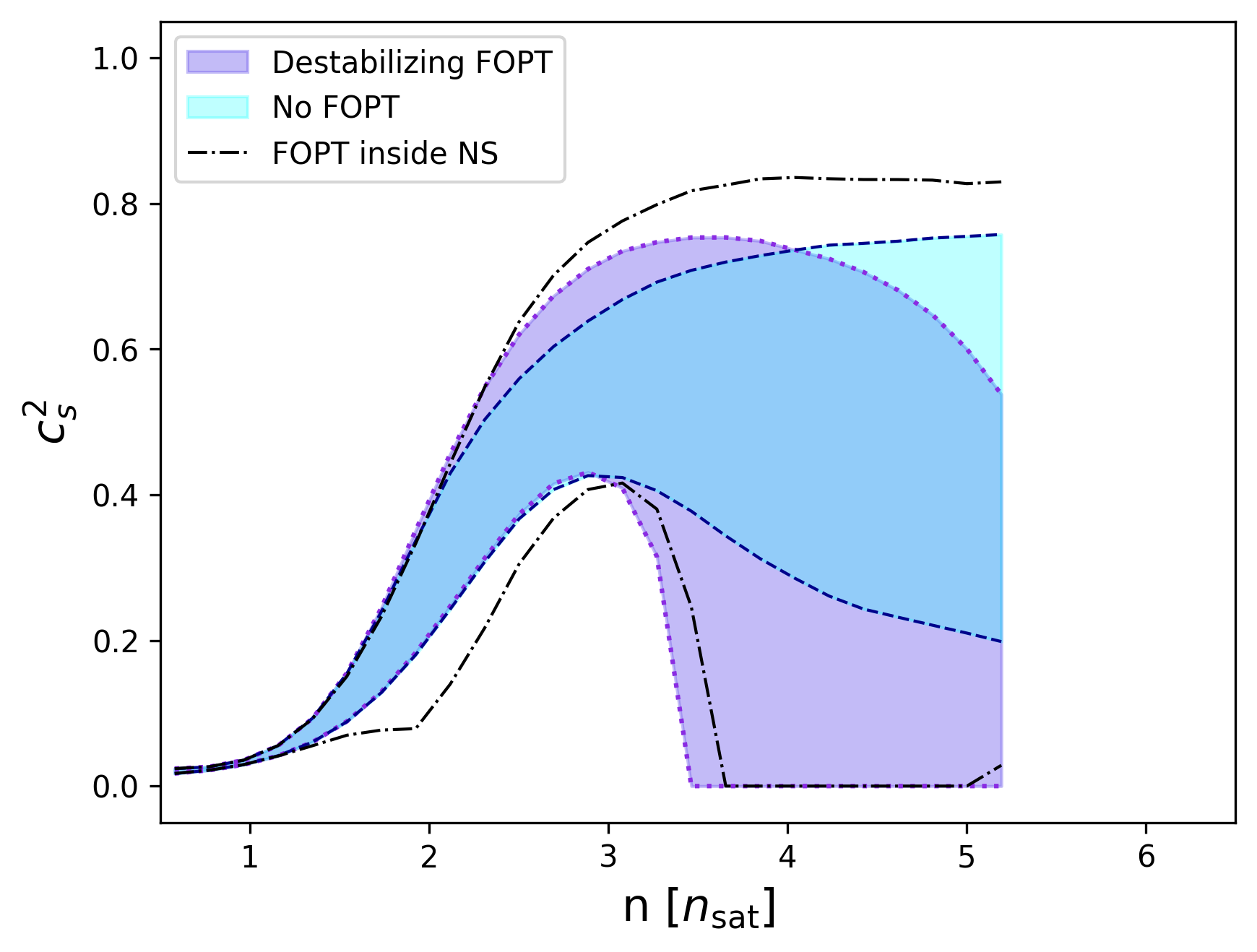}
    \caption{68\%-CI for $c^2_s$ for different sets for less aggressive inputs: cEFT up to 1.1$n_{\rm sat}$, conservative QCD input, NICER PSR J0740+662, radio measurement PSR J0348+0432 and GW170818 data. The plot shows the conditional posterior distribution $\mathrm{P}(c^2_s|n, n<\ntov)$.}
    \label{fig:cs2_conserv}
\end{figure}

In this supplemental material I compare the results obtained in the main text using state-of-the-art inputs with less aggressive choices: conservative QCD input, cEFT constraints up to 1.1 n$_{\rm sat}$, only NICER PSR J0740+662 out of all x-ray observations, along with radio and GW170817 data.

Confidence intervals for the sounds speed are shown in \cref{fig:cs2_conserv}. The most noticeable difference to \cref{cs2} upper left subplot is the lack of forced softening at the highest densities reached in NSs for \textit{No FOPT} and \textit{PT inside NS} sets. This happens due to the change from marginalized QCD to conservative QCD input. While \textit{destabilizing FOPT} has a manually imposed softening to $c^2_s = 0$ at the end of the stable branch, the peak in sound speed reaches lower values. This is because, as explained in Section IIIA, dramatic softening requires a stiffer EoS when utilizing more information from the pQCD limit. 

It is important to note that, according to \cite{Komoltsev:2023zor}, using conservative QCD input, a stiff EoS requires extreme behavior above $n_{\text{TOV}}$, such as a large PT of $\Delta n \sim 20 n_{\rm sat}$ followed by a segment of $c^2_s \approx 1$ to remain consistent with pQCD. The marginalized QCD likelihood function penalizes such behavior.

Another noticeable difference is the wider 68\%-CI at the intermediate densities below $n < 3n_{\rm sat}$ for the less aggressive inputs. This is due to the number of x-ray observations used. While mass constraints from radio measurements and NICER PSR J0740 require stiffening of EoS at intermediate densities, the 12 x-ray observations impose much stronger constraints, necessitating greater stiffening and narrowing the band.

\section{Acknowledgments}

The author expresses gratitude to Aleksi Kurkela for providing thorough supervision throughout the development of the manuscript and for suggesting the topic. The author thanks Tyler Gorda, Joonas N\"attil\"a, Aleksi Vuorinen, Jacquelyn Noronha-Hostler, Philippe Landry and Reed Essick for their valuable discussions and insightful comments.

\newpage

\bibliography{main.bib}

\end{document}